\documentclass[a4paper,12pt]{article}

\usepackage[T1]{fontenc}
\usepackage{times}
\usepackage[colorlinks=true,linkcolor=black,citecolor=black,urlcolor=blue]{hyperref}

\usepackage{geometry}
\usepackage{url}
\usepackage{comment}
\usepackage{mathtools}
\usepackage{algorithm2e}
\usepackage{amssymb}
\usepackage{csquotes}
\usepackage{xcolor}
\usepackage{amsmath}
\usepackage{amssymb}
\usepackage{graphicx}
\usepackage{float}
\usepackage{csquotes}
\usepackage{csquotes}
\usepackage{url}
\usepackage{graphicx}
\usepackage{booktabs}
\usepackage{siunitx}
\usepackage{titlesec}
\usepackage{}
\titleformat{\section}
{\bfseries\uppercase}{\thesection.}{1em}{}
\titleformat{\subsection}
{\bfseries}{\thesection.\thesubsection.}{1em}{}

\usepackage{graphicx} 
\usepackage{multirow} 
\usepackage{cite}
\usepackage{breakurl}
\usepackage{indentfirst}
\usepackage{amsmath, amssymb, amsfonts, bm}
\usepackage{txfonts}
\usepackage{enumitem}
\usepackage{xcolor}
\usepackage{enumitem}

\titlespacing*{\subsection}{0pt}{1.5em}{0.2em}
\titlespacing*{\section}{0pt}{1.5em}{0.2em}

\renewcommand\eqref[1]{Equation~\ref{#1}}

\renewcommand{\thesection}{\arabic{section}}
\renewcommand{\thesubsection}{\arabic{subsection}}

\makeatletter
\renewcommand\@biblabel[1]{#1.}
\makeatother

\newlength{\bibitemsep}
\newlength{\bibparskip}
\let\oldthebibliography\thebibliography
\renewcommand\thebibliography[1]{%
  \oldthebibliography{#1}%
  \setlength{\parskip}{\bibitemsep}%
  \setlength{\itemsep}{\bibparskip}%
}
 
\begin{document}
\begin{center}
	\includegraphics[width=3.50in]{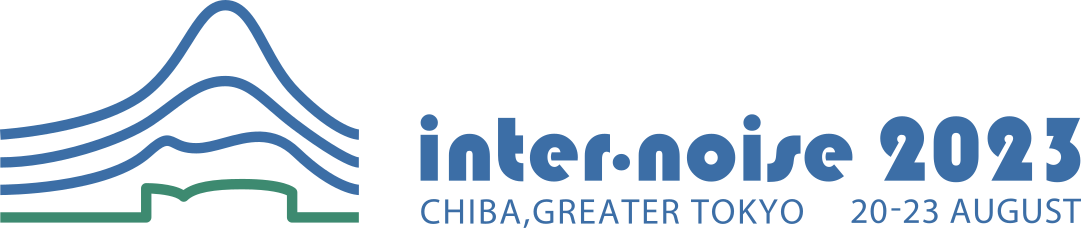}
\end{center}
\vskip.5cm

\begin{flushleft}
\fontsize{16}{20}\selectfont\bfseries
\textcolor{red}{E-PANNs: Sound Recognition Using Efficient Pre-trained Audio Neural Networks} \\
\end{flushleft}
\vskip0.5cm


    
\begin{flushleft}
Arshdeep Singh\footnote{arshdeep.singh@surrey.ac.uk}, Haohe Liu\footnote{haohe.liu@surrey.ac.uk}, Mark D. Plumbley\footnote{ m.plumbley@surrey.ac.uk}\\
Centre for Vision, Speech and Signal Processing (CVSSP),\\
University of Surrey, UK 
\end{flushleft}
\renewcommand\abstractname{\textsc{Abstract}}


\hspace{-1.5cm}\textbf{\centerline{ABSTRACT}}
\noindent
\textit{Sounds carry an abundance of information about activities and events in our everyday environment, such as traffic noise, road works, music, or people talking. Recent machine learning methods, such as convolutional neural networks (CNNs), have been shown to be able to automatically recognize sound activities, a task known as audio tagging. One such method, pre-trained audio neural networks (PANNs), provides a neural network which has been pre-trained on over 500 sound classes from the publicly available AudioSet dataset, and can be used as a baseline or starting point for other tasks. However, the existing PANNs model has a high computational complexity and large storage requirement. This could limit the potential for deploying PANNs on resource-constrained devices, such as on-the-edge sound sensors, and could lead to high energy consumption if many such devices were deployed. In this paper, we reduce the computational complexity and memory requirement of the PANNs model by taking a pruning approach to eliminate redundant parameters from the PANNs model. The resulting Efficient PANNs (E-PANNs) model, which requires 36\% less computations and 70\% less memory, also slightly improves the sound recognition (audio tagging) performance. The code for the E-PANNs model  has been released under an open source license.}

\section{INTRODUCTION}
\noindent
Everyday sound environments include a wide range of sound activities and events, such as traffic  noise, road works, key jangling, music, coughing or people talking. These environmental sound  activities contain an abundance of information and can potentially be used in various applications
including public security surveillance, monitoring activities in a home for assisted living, healthcare, and improving the office, workplace and urban environment.

\begin{figure}[t]
    \centering
    \includegraphics[scale=0.3]{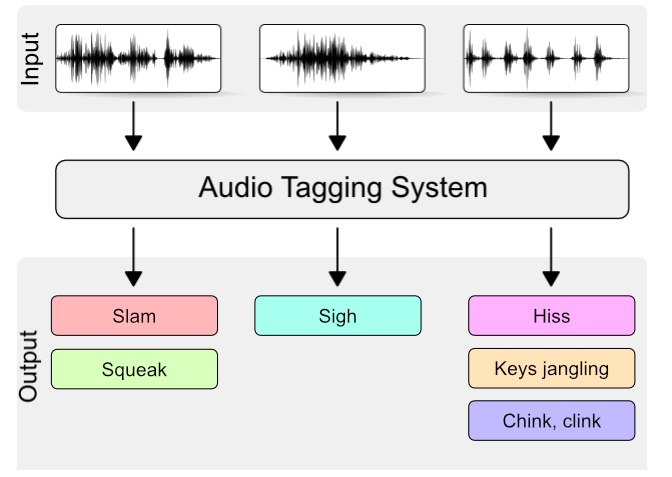}
    \caption{An audio tagging system recognising various sound activities in a given audio recording \cite{Fonseca2019audio}.}
    \label{fig: audio tagging system}
\end{figure}

Recent advances in machine learning infrastructure and the  availability of large-scale dataset such as AudioSet \cite{gemmeke2017audio} have attracted artificial intelligence and machine learning (AI/ML) researchers to develop methods for automatic sound activity recognition, commonly known as audio tagging.  A typical audio tagging system is shown in Figure \ref{fig: audio tagging system}. It takes audio recordings from the surrounding  using microphones and then, recognises various sound activities that occur in the surrounding area. Audio tagging systems using convolutional neural networks (CNNs) have shown promising performance compared to the traditional hand-crafted methods \cite{kong2020panns}. However, CNNs are resource-hungry due to the high computational cost arising from multiply-accumulate operations (MACs) and from the memory requirement for CNNs. For example, one of the best performing  audio tagging networks from  the pre-trained audio neural networks (PANNs) \cite{kong2020panns} framework has approximately 81M parameters and requires more than 50G MACs for inference corresponding to a 10s audio clip. Due to this,  it may be challenging to deploy such large-scale CNNs on resource-constrained devices having a limited power budget and limited memory, such as smart phones or internet of things (IoT) devices.  Moreover,  when large-scale CNNs are used as a feature extractor or as a classifier for other downstream tasks such as acoustic scene classification \cite{martin2021low}, the high computational cost of the CNNs makes them slow during inference and particularly  
during the training process, where the large-scale CNNs may consume more energy and emit more $\text{CO}_2$. For instance, an NVIDIA RTX-2080 Ti GPU used to train machine learning models for 48 hours generates the equivalent amount of $\text{CO}_2$ to that  emitted by an average car driven for 13 miles\footnote{Machine learning $\text{CO}_2$ estimator: \url{https://mlco2.github.io/impact/##compute}}.   
Therefore, large-scale CNNs are not efficient and environmental-friendly  despite performing well.


\noindent\textbf{Our contributions:} To reduce computational complexity and memory storage of large-scale CNNs such as PANNs,
we eliminate redundant parameters and reduce computational complexity from the PANNs model by taking a pruning approach.   The resulting Efficient PANNs (E-PANNs) model has significantly less MACs and memory storage  with slightly improved performance  compared to that of the original PANNs model. We make a real-time sound recognition demonstration using E-PANNs publicly available\footnote{\url{https://github.com/Arshdeep-Singh-Boparai/E-PANNs.git}}. 

The rest of the paper is organised as follows. Section \ref{sec: background} introduces some background including a brief overview of convolutional neural networks,  and background on existing audio tagging systems and  the PANNs architecture. Section \ref{sec: Proposed metho} presents the proposed methodology used to obtain E-PANNs.  Next, the experimental setup and dataset used for experiments is explained in Section \ref{sec: Experimental setup}. Section \ref{sec: performance analysis} presents experimental analysis. Finally, Section \ref{sec: Conclusion} concludes the paper.

\section{Background}
\label{sec: background}


\subsection{Convolutional Neural Networks (CNNs)}
\label{sec: CNNs}
\begin{figure}[h]
    \centering
    \includegraphics[scale = 0.68]{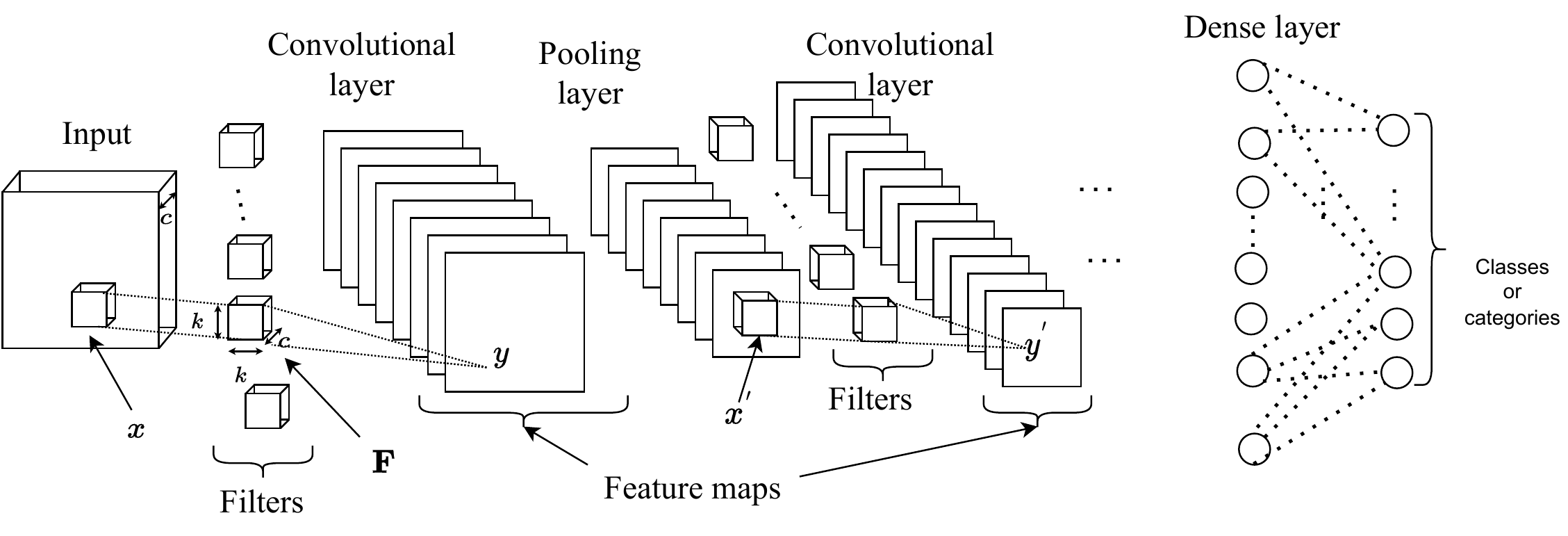}
    \caption{Convolutional neural network (CNN) architecture comprising of convolutional, pooling and dense layer.}
    \label{fig: CNN architecture}
\end{figure}

\noindent Convolutional neural networks (CNNs) are a type of artificial neural network inspired by biological nervous systems such as the human brain. CNNs are designed to learn from examples or from a dataset for an underlying task such as classification of sound activity in the surrounding area. They learn through an optimization process to update their parameters including weights and filters across various  types of intermediate layers, such as convolutional, pooling, and dense layers. An architecture of a CNN  is shown in Figure \ref{fig: CNN architecture}. A convolutional layer has multiple feature maps, where each feature map is produced by the convolutional operation on input and  a filter.

In a CNN, filters are small matrices of size $(k \times k)$ with $c$ channels that are convolved across the input data. The convolution operation, as given in Equation \ref{Equ: convo operation},  is a multiply-accumulate operation (MAC) that involves sliding the filter, $\mathbf{F}$, over the input, performing element-wise multiplications between the filter and the corresponding input patch $x$ of size ($c \times k \times k$), and summing up the results to produce a single value, $y$. This process is repeated for every possible position of the filter across the input, resulting in a \textit{feature map}.  Similarly, other feature maps are produced using other filters in a given convolutional layer. Subsequently a bias $b$ and a non-linear activation function $f(.)$ is applied to the elements of feature maps.  Other than convolutional, pooling and dense layers, some CNNs might  have residual  blocks, which contain skip or shortcut connections that bypass one or more of the  convolutional layers.  For an introduction to various architectures of CNNs, see the article \cite{gu2018recent}.

\begin{equation}
    y = f(\sum_{i=1} ^{c}\sum_{k_1=1}^{k} \sum_{k_2=1}^{k}  (x_{i,k_1,k_2} \times \mathbf{F}_{i, k_1, k_2}) + b) .
    \label{Equ: convo operation}
\end{equation}


\subsection{Existing Audio Tagging Frameworks}
\noindent With the release of the large-scale AudioSet dataset \cite{gemmeke2017audio}, which has  2M audio examples and 527 classes, several researchers have conducted studies to improve the performance of neural networks on AudioSet classification and in particular CNNs. Methods for AudioSet classification  include CNNs  \cite{kong2020panns}, CNNs with residual blocks \cite{verbitskiy2022eranns} and and self-attention based Transformers \cite{gong2022contrastive}. A summary of the existing methods alongwith their performance and the number of parameters is given in Table \ref{tab: panns papers}.  While Transformer-based methods perform better than CNNs,  Transformers have high computational complexity, which makes their deployment on low-powered devices difficult compared to the CNNs. We shall therefore focus on CNNs in this paper.




\begin{table}[h]
	\caption{\color{black}{Mean average precision (mAPs) obtained for AudioSet evaluation dataset using various audio tagging systems with their number of parameters.}}
	\centering
	\resizebox{0.68\textwidth}{!}{
		\begin{tabular}{SSSSS} \toprule
            {Method} & {Neural Network type} & {Parameters (x $10^6$)}  & {mAPs} \\ \midrule
   			{PANNs-CNN6 \cite{kong2020panns}} & {Plain} & {4.8} & {0.343}   \\
			{PANNs-CNN10  \cite{kong2020panns}} & {Plain} & {5.2} & {0.380}  \\
            {ERANNs-2-5 \cite{verbitskiy2022eranns}} & {Residual} & {38.2} & {0.446}   \\
			{ERANNS-1-6 \cite{verbitskiy2022eranns}} & {Residual} & {54.5} & {0.450}  \\
        	{PANNs (ResNet38) \cite{kong2020panns}} & {Residual} & {73.78} & {0.434}   \\
			{PANNs-CNN14  \cite{kong2020panns}} & {Plain} & {80.75}  & {0.431}  \\
			{PANNs (Wavegram-Logmel-CNN)  \cite{kong2020panns}} & {Plain} & {81.06}  & {0.439}  \\
			{AST \cite{gong21b_interspeech}} & {Transformer} & {88.10}  & {0.459}  \\
     		{AST(ensemble)  \cite{gong21b_interspeech}} & {Transformer} & {526.6} & {0.485}   \\
   \bottomrule
 	\end{tabular}}
	\label{tab: panns papers}
\end{table}

\subsection{PANNs-CNN14 Architecture for AudioSet Classification}

\noindent The primary motivation for the development of PANNs was to provide pre-training systems for audio pattern recognition on extensive datasets, in this case, the AudioSet dataset \cite{gemmeke2017audio}, which can be used as a baseline network to extract features or for a classification task to other audio-related tasks. The authors proposed several architectures for PANNs, including CNN-14, which  demonstrated promising performance on various audio pattern recognition tasks. Experiments demonstrated that PANNs can generalize well to other tasks with limited training data and outperform models trained from scratch on those tasks.

The architecture of the PANNs-CNN14 model, including the number of parameters across each convolutional layers, the total number of parameters and the model size is shown in Figure \ref{fig: architecutre CNN14} (a) in the  Appendix. It  has  six convolutional blocks, each with two convolutional layers, followed by a batch normalization layer and a ReLU activation function. The number of filters in each convolutional block gradually increases from 64 to 2048 from layer to layer. Finally, there is a dense layer with a sigmoid activation function that outputs the predicted probabilities for each class. 

The PANNs-CNN14 model takes a log-mel spectrogram of size (1000 $\times$ 64) computed from the  10-second-length audio input. The model is trained with data augmentation techniques such as Mixup \cite{zhang2017mixup} and SpecAugment \cite{park2019specaugment} for 600k iterations. For more details on CNN14 and other PANNs models, see \cite{kong2020panns}.




\section{Proposed Methodology to Reduce Complexity}
\label{sec: Proposed metho}
\noindent To reduce the computational complexity and memory storage of CNNs such as PANNs-CNN14, we use a filter pruning approach \cite{liang2021pruning, luo2017thinet, ding2019compressing} that involves removing or \enquote{pruning} unimportant filters from the network, i.e. these filters that contribute less to the performance of the CNN. Filter pruning is inspired by the idea that some filters in a CNN are redundant or have a negligible contribution to the overall accuracy of the network \cite{frankle2018lottery, denil2013predicting}. These filters can be safely removed from the network without significantly affecting its performance. For example, previous studies  \cite{singh2019deep, singh2020svd} found that 73\% of the filters in a SoundNet network \cite{aytar2016soundnet} do not provide discriminative information across different acoustic scene classes, and eliminating such filters give similar performance to that of using all the original filters in SoundNet.  For a survey on pruning techniques, see \cite{liang2021pruning}.

\noindent The typical  steps in a filter pruning process are:

\begin{enumerate}
    \item  For a baseline, train the original network to a desired level of accuracy, or utilise an already existing pre-trained network;
    \item Rank the filters based on a certain \enquote{importance} criterion, such as their relevance in contributing to the performance of the network;
    \item Remove the least important filters and  their corresponding  feature maps  from the network to obtain a pruned network;
    \item Fine-tune the pruned network  to recover some of the lost accuracy.
\end{enumerate}

There are several benefits of filter pruning in CNNs. Firstly, it can significantly reduce the size of the network, reducing the memory footprint and computation time required for inference by removing filters and the corresponding feature maps generated by the filters.  Secondly, the robustness of the network in maintaining performance and generalization capabilities can improve by removing filters that may be sensitive to small perturbations in the input data.

In this paper, we will apply the filter pruning approach to the PANNs-CNN14 model, to produce an efficient PANNs model, E-PANNs. To obtain E-PANNs, we follow a three step pipeline as shown in Figure  \ref{fig: flow diagram}. A detailed description of the steps is given below, 
\\

\noindent \textbf{(Step 1) Take a baseline PANNs network}: We take  the publicly available pre-trained CNN14 from PANNs and denote it as PANNs-CNN14. PANNs-CNN14 has approximately 81M parameters and requires 21G MACs\footnote{MACs computation Pytorch package.: \url{https://pypi.org/project/thop/}} for inference of a 10s audio clip with a performance metric, mAPs of 0.431. \\

\noindent \textbf{(Step 2) Compute filter importance across convolutional layers of the baseline network:}
We compute the relevance of the filters to decide whether to retain or prune  the filter for each layer of PANNs-CNN14 independently  using  (a), (b), and (c) steps as shown in Figure \ref{fig: flow diagram}.

Assume that there are $n$ filters in a given convolutional layer. With these $n$ filters, first, we measure how well a given filter produces the  output in the convolutional layer. Our hypothesis is that a filter producing significant output can capture specific patterns or features in the input significantly, which would be useful for the subsequent convolutional layers of CNN to better understand the input data at different levels of abstraction. Therefore, we consider a filter producing significant output more important than others.
Then, we measure the importance of each filter and rank the filters according to their importance in a given convolutional layer.  Subsequently, the process (a)-(c) is repeated for other convolutional layers as well. More information  about the importance calculation of the filters  can be found in \cite{singh2023efficient}. \\

\noindent \textbf{(Step 3) Obtaining E-PANNs:} Once we obtain the importance of each filter in various convolutional layers in PANNs-CNN14, we eliminate $p$  least important filters and their corresponding feature maps from various convolutional layers. This results in a pruned network. The pruned network is then re-trained or fine-tuned with same data as used for training the baseline PANNs-CNN14 network, to regain some of the lost performance owing to elimination of the filters and their corresponding feature map. Finally, an efficient version of PANNs-CNN14, E-PANNs is obtained. 

\begin{figure}[t]
    \centering
    \includegraphics[scale=0.5]{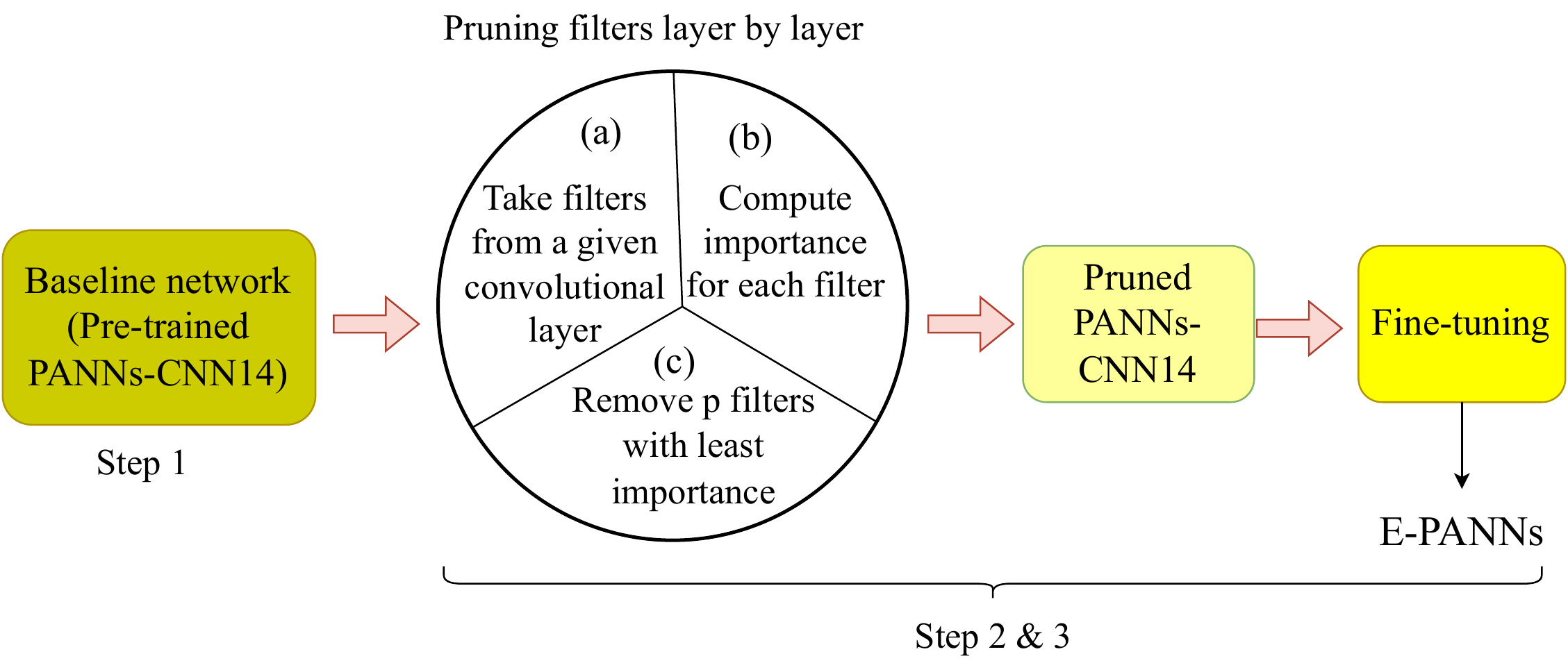}
    \caption{A flow diagram to obtain E-PANNs.}
    \label{fig: flow diagram}
\end{figure}

\section{Experimental Setup}
\label{sec: Experimental setup}

\noindent
In this section, we describe the experimental setup to prune the PANNs-CNN14 and then fine-tune the pruned network. 

In PANNs-CNN14 model, the last six convolutional layers (C7 to C12) yield approximately 99\% of the total network parameters (See architecture in Figure \ref{fig: architecutre CNN14} (a) in the Appendix), therefore we prune  $p \in \{25\%, 50\%, 75\%\}$ filters from the last six convolutional layers. The fine-tuning process of the pruned PANNs-CNN14 is performed in the same way as  that used to train the original PANNs-CNN14 \cite{kong2020panns}. The  optimization procedure uses the same loss function, data augmentation and batch size etc, but with fewer iterations, which are approximately 450k for the pruned PANNs-CNN14. To evaluate the performance of the proposed E-PANNs, we test it on the  audio tagging problem using  the AudioSet evaluation dataset.

\section{Performance Analysis}
\label{sec: performance analysis}

\begin{figure}[h]
    \centering
    \includegraphics[scale=0.4]{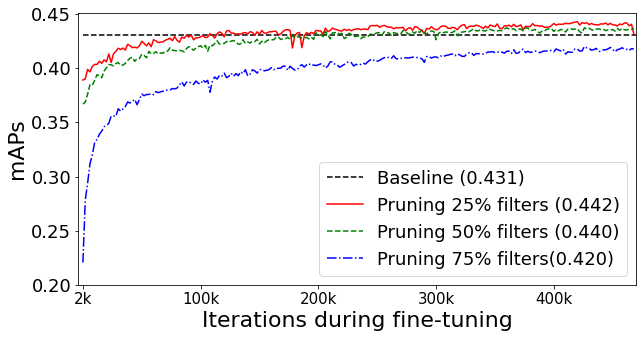}
    \caption{Convergence during fine-tuning process of E-PANNs when different number of filters are pruned from the baseline network. In brackets, the maximum mAPs obtained is also given.}
    \label{fig: pruning ratio vs convergence}
\end{figure}

\noindent
Figure \ref{fig: pruning ratio vs convergence} shows a convergence plot obtained during  the fine-tuning process of  E-PANNs, when different numbers of filters are pruned from the baseline network. We find that E-PANNs  recovers the same performance as that of the baseline network, even after pruning 25\% and 50\% filters, with only 100k and 200k iterations respectively. We also find that it uses at least 3 times fewer iterations compared to the baseline network to achieve  mAPs equal to 0.431.  Therefore, E-PANNs can be used as an alternative to the original PANNs with an advantage of less computation and memory requirement. 



\begin{figure}[h]
    \centering
    \includegraphics[scale=0.47]{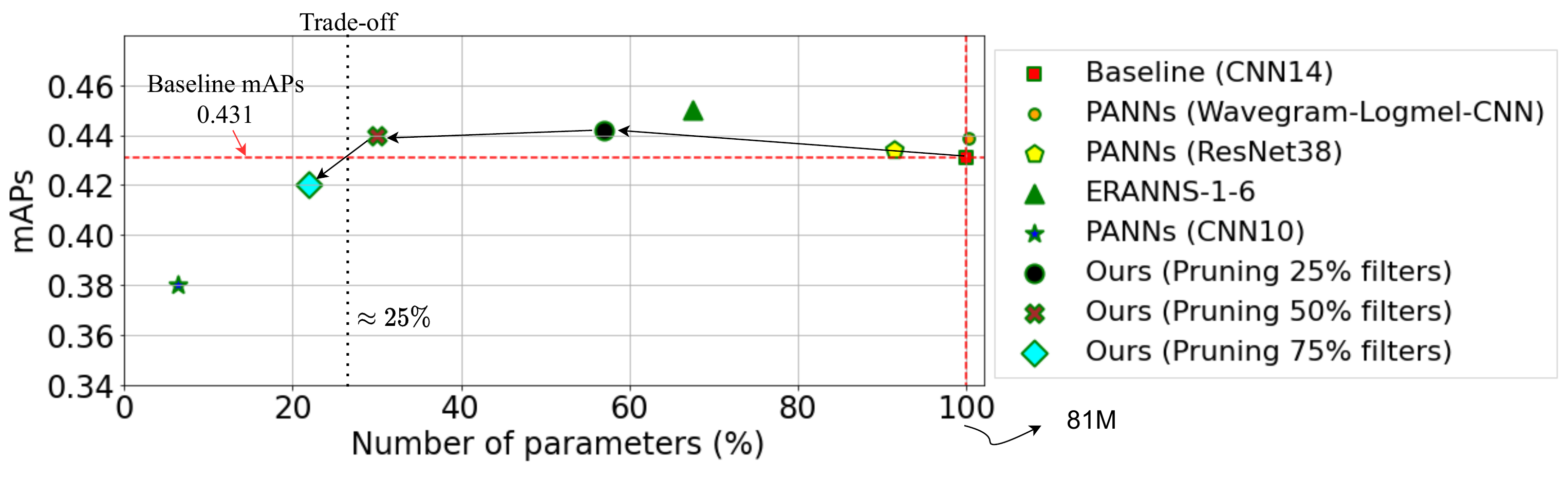}
    \caption{mAPs and the number of parameters obtained using the proposed method and the other existing CNN-based methods.}
    \label{fig: exsiting methods comparisons}
\end{figure}

We find that pruning 25\% of the  filters across C7 to C12 layers of the baseline network removes 41\% of the parameters  and 24\% of the  MACs with an improved  mAPs equal to 0.442,  compared to that of the baseline network. Also after pruning 50\% of the filters, the mAPs obtained are better than that of the baseline network, with 70\% fewer parameters and 36\% fewer MACs. The architecture of the E-PANNs network obtained after pruning 50\% of the filters is shown in Figure \ref{fig: architecutre CNN14} (b) in the Appendix. On the other hand, when  75\% of the filters are removed, the mAPs is 0.420 with 78\% fewer parameters and 46\% fewer MACs. This suggests that there is a trade-off between the number of parameters pruned and the mAPs. As shown in Figure \ref{fig: exsiting methods comparisons} (see arrow trend), we find that when the number of parameters in E-PANNs is less than 25\% of the total number of parameters in the  baseline network, and that the mAPs obtained  using E-PANNs becomes less than the mAPs obtained using the baseline network. Therefore, the maximum number of parameters that can be pruned from the baseline network, PANNs-CNN14, without any degradation in performance are  approximately 75\%.

Figure \ref{fig: exsiting methods comparisons} also shows the number of parameters and mAPs obtained using the other existing CNN-based methods as given in Table \ref{tab: panns papers}.  Our proposed method gives better mAPs  with 70\% fewer parameters and 60\% fewer parameters respectively compared to the best performing PANNs-(Wavegram-Logmel-CNN) and the PANNs-ResNet38 network. In general,  compared to other methods such as PANNs-CNN10 or ERANNs-1-6 \cite{verbitskiy2022eranns}, we find that there is a trade-off between the number of parameters and the mAPs. 

\section{Conclusion}
\label{sec: Conclusion}
\noindent Convolutional neural networks (CNNs) are effective for audio recognition, but can be costly. We make these CNNs more efficient by reducing their computational cost and memory storage. We present a framework to obtain E-PANNs from a large-scale pre-trained audio neural network (PANNs) for audio tagging. We find that removing few of the unimportant filters from original PANNs reduces the computational complexity by 36\% and 70\% fewer parameters with an improved performance. Therefore, E-PANNs can be used as an alternative to the original PANNs for the downstream tasks for feature extraction or classification due to the less computational complexity and low memory requirement of E-PANNs compared to that of the PANNs.

\section{Acknowledgements}

\noindent This work was partly supported by Engineering and Physical Sciences Research Council (EPSRC) Grant EP/T019751/1 \enquote{AI for Sound (AI4S)}. For the purpose of open access, the authors have applied a Creative Commons Attribution (CC BY) licence to any Author Accepted Manuscript version arising.

\bibliographystyle{unsrt}
\bibliography{sample} 

\begin{thebibliography}{10}

\bibitem{gemmeke2017audio}
Jort~F Gemmeke, Daniel~PW Ellis, Dylan Freedman, Aren Jansen, Wade Lawrence,
  R~Channing Moore, Manoj Plakal, and Marvin Ritter.
\newblock Audio set: An ontology and human-labeled dataset for audio events.
\newblock In {\em IEEE International Conference on Acoustics, Speech and Signal
  Processing (ICASSP)}, pages 776--780. IEEE, 2017.

\bibitem{kong2020panns}
Qiuqiang Kong, Yin Cao, Turab Iqbal, Yuxuan Wang, Wenwu Wang, and Mark~D
  Plumbley.
\newblock {PANNs}: Large-scale pretrained audio neural networks for audio
  pattern recognition.
\newblock {\em IEEE/ACM Transactions on Audio, Speech, and Language
  Processing}, 28:2880--2894, 2020.

\bibitem{martin2021low}
Irene Mart{\'\i}n-Morat{\'o}, Toni Heittola, Annamaria Mesaros, and Tuomas
  Virtanen.
\newblock {Low-complexity acoustic scene classification for multi-device audio:
  Analysis of DCASE 2021 Challenge systems}.
\newblock {\em DCASE Workshop}, 2021.

\bibitem{Fonseca2019audio}
Eduardo Fonseca, Manoj Plakal, Frederic Font, Daniel P.~W. Ellis, and Xavier
  Serra.
\newblock Audio tagging with noisy labels and minimal supervision.
\newblock In {\em DCASE2019 Workshop}, NY, USA, 2019.

\bibitem{gu2018recent}
Jiuxiang Gu, Zhenhua Wang, Jason Kuen, Lianyang Ma, Amir Shahroudy, Bing Shuai,
  Ting Liu, Xingxing Wang, Gang Wang, Jianfei Cai, et~al.
\newblock {Recent advances in convolutional neural networks}.
\newblock In {\em Pattern Recognition}, volume~77, pages 354--377. Elsevier,
  2018.

\bibitem{verbitskiy2022eranns}
Sergey Verbitskiy, Vladimir Berikov, and Viacheslav Vyshegorodtsev.
\newblock {ERANNS}: Efficient residual audio neural networks for audio pattern
  recognition.
\newblock {\em Pattern Recognition Letters}, 161:38--44, 2022.

\bibitem{gong2022contrastive}
Yuan Gong, Andrew Rouditchenko, Alexander~H Liu, David Harwath, Leonid
  Karlinsky, Hilde Kuehne, and James~R Glass.
\newblock Contrastive audio-visual masked autoencoder.
\newblock In {\em International Conference on Learning Representations}, 2023.

\bibitem{gong21b_interspeech}
Yuan Gong, Yu-An Chung, and James Glass.
\newblock {AST: Audio Spectrogram Transformer}.
\newblock In {\em Proc. Interspeech 2021}, pages 571--575, 2021.

\bibitem{zhang2017mixup}
Hongyi Zhang, Moustapha Cisse, Yann~N Dauphin, and David Lopez-Paz.
\newblock mixup: Beyond empirical risk minimization.
\newblock {\em International Conference on Learning Representations (ICLR)},
  2017.

\bibitem{park2019specaugment}
Daniel~S Park, William Chan, Yu~Zhang, Chung-Cheng Chiu, Barret Zoph, Ekin~D
  Cubuk, and Quoc~V Le.
\newblock {SpecAugment}: A simple data augmentation method for automatic speech
  recognition.
\newblock {\em Interspeech}, 2019.

\bibitem{liang2021pruning}
Tailin Liang, John Glossner, Lei Wang, Shaobo Shi, and Xiaotong Zhang.
\newblock Pruning and quantization for deep neural network acceleration: A
  survey.
\newblock {\em Neurocomputing}, 461:370--403, 2021.

\bibitem{luo2017thinet}
Jian-Hao Luo, Jianxin Wu, and Weiyao Lin.
\newblock {ThiNet: A filter level pruning method for deep neural network
  compression}.
\newblock In {\em Proceedings of the IEEE International Conference on Computer
  Vision}, pages 5058--5066, 2017.

\bibitem{ding2019compressing}
Haisong Ding, Kai Chen, and Qiang Huo.
\newblock Compressing {CNN-DBLSTM} models for {OCR} with teacher-student
  learning and tucker decomposition.
\newblock In {\em Pattern Recognition}, volume~96, page 106957. Elsevier, 2019.

\bibitem{frankle2018lottery}
Jonathan Frankle and Michael Carbin.
\newblock The lottery ticket hypothesis: Finding sparse, trainable neural
  networks.
\newblock In {\em International Conference on Learning Representations}, 2019.

\bibitem{denil2013predicting}
Misha Denil, Babak Shakibi, Laurent Dinh, M.A. Ranzato, and Nando De~Freitas.
\newblock Predicting parameters in deep learning.
\newblock In {\em Advances in Neural Information Processing Systems}, pages
  2148--2156, 2013.

\bibitem{singh2019deep}
Arshdeep Singh, Padmanabhan Rajan, and Arnav Bhavsar.
\newblock Deep hidden analysis: A statistical framework to prune feature maps.
\newblock In {\em IEEE International Conference on Acoustics, Speech and Signal
  Processing (ICASSP)}, pages 820--824. IEEE, 2019.

\bibitem{singh2020svd}
Arshdeep Singh, Padmanabhan Rajan, and Arnav Bhavsar.
\newblock {SVD-based redundancy removal in 1-D CNNs for acoustic scene
  classification}.
\newblock In {\em Pattern Recognition Letters}, volume 131, pages 383--389.
  Elsevier, 2020.

\bibitem{aytar2016soundnet}
Yusuf Aytar, Carl Vondrick, and Antonio Torralba.
\newblock {SoundNet: Learning sound representations from unlabeled video}.
\newblock {\em Advances in Neural Information Processing Systems}, 29, 2016.

\bibitem{singh2023efficient}
Arshdeep Singh and Mark~D Plumbley.
\newblock {Efficient CNNs via Passive Filter Pruning}.
\newblock {\em arXiv preprint arXiv:2304.02319}, 2023.

\end{thebibliography}

\newpage
\section*{Appendix}

\textbf{Architecture of baseline network: PANNs-CNN14 model and efficient PANNs-CNN14.}

\label{sec: E-PANNs architecture}
\begin{figure}[h]
    \centering
    \includegraphics[scale=0.72]{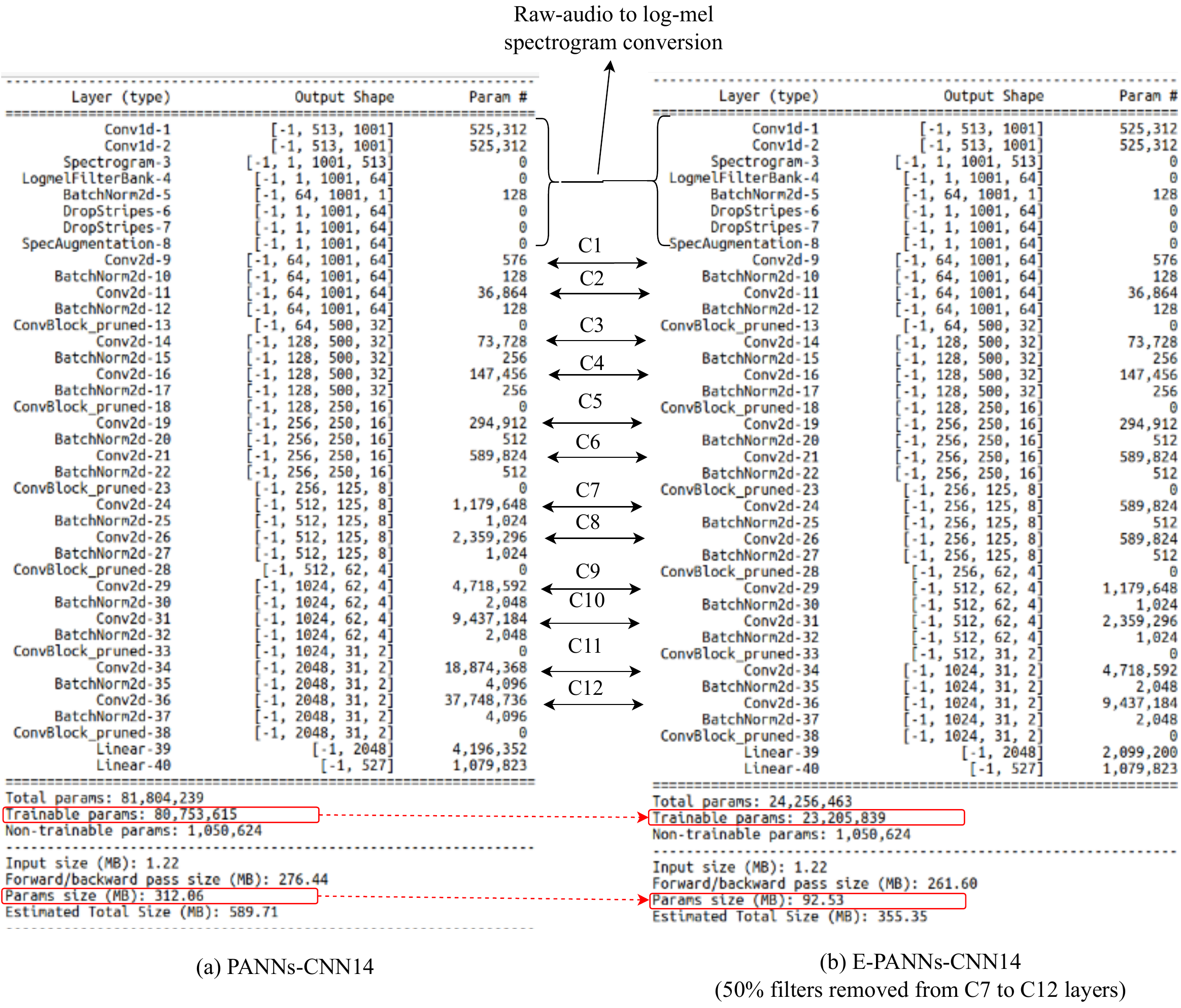}
    \caption{(a) The architecture of baseline network PANNs-CNN14. (b) The architecture of E-PANNs obtained after removing 50\% of the filters from C7 to C12 convolutional layers of  PANNs-CNN14. The number of parameters for each convolutional layer and the parameter (model) size is also shown. The PANNs-CNN14 has 12 trainable convolutional layers, denoted as C1 to C12. The first two convolutional layers (conv1d-1. conv1d-2) are non-trainable and their purpose is to convert the raw audio signal to log-mel spectrogram.}
    \label{fig: architecutre CNN14}
\end{figure}

\end{document}